\documentclass[preprint,showpacs,preprintnumbers,amsmath,amssymb]{revtex4}

\usepackage{graphicx}
\usepackage{dcolumn}
\usepackage{bm}

\begin{document}


 \title{Superconductivity in Geometrically Frustrated Pyrochlore RbOs$_2$O$_6$}

\author{M.~Br\"uhwiler}
 \email{bruehwiler@solid.phys.ethz.ch}
 \affiliation{Laboratory for Solid State Physics, ETH Z\"urich, 8093 Z\"urich, Switzerland.}
\author{S.M.~Kazakov}%
 \affiliation{Laboratory for Solid State Physics, ETH Z\"urich, 8093 Z\"urich, Switzerland.}
\author{N.D.~Zhigadlo}%
 \affiliation{Laboratory for Solid State Physics, ETH Z\"urich, 8093 Z\"urich, Switzerland.}
\author{J.~Karpinski}%
 \affiliation{Laboratory for Solid State Physics, ETH Z\"urich, 8093 Z\"urich, Switzerland.}
\author{B.~Batlogg}%
 \affiliation{Laboratory for Solid State Physics, ETH Z\"urich, 8093 Z\"urich, Switzerland.}

\date{\today}

\newcommand{\fu}{$\mathrm{RbOs_2O_6}$}
\newcommand*{\unit}[1]{\,\mathrm{#1}}

\begin{abstract}
We report the basic thermodynamic properties of the new
geometrically frustrated $\beta$-pyrochlore bulk superconductor
\fu\ with a critical temperature $T_\mathrm{c} = 6.4\unit{K}$.
Specific heat measurements are performed in magnetic fields up to
$12\unit{T}$. The electronic density of states at the Fermi level
in the normal state results in $\gamma =  (33.7 \pm
0.2)\unit{mJ/mol_{\mathrm{f.u.}}/K^2}$. In the superconducting
state, the specific heat follows conventional BCS-type behavior
down to $1\unit{K}$, i.e.~over three orders of magnitude in
specific heat data. The upper critical field slope at
$T_\mathrm{c}$ is $1.2\unit{T/K}$, corresponding to a
Maki-parameter $\alpha = 0.64 \pm 0.1$. From the upper critical
field $\mu_0 H_{\mathrm{c2}} \approx 6\unit{T}$ at $0\unit{K}$, we
estimate a Ginzburg-Landau coherence length $\xi \approx
74\unit{\AA}$. \fu\ is the second reported metallic AB$_2$O$_6$
type pyrochlore compound after $\mathrm{KOs_2O_6}$, and one of
only three pyrochlore superconductors in addition to
Cd$_2$Re$_2$O$_7$ and $\mathrm{KOs_2O_6}$.
\end{abstract}

\maketitle


Nonconducting magnetic compounds with d electrons on a
3-dimensional triangular lattice have been intensely studied. One
of the magnetic ground states evolving out of the geometrically
frustrated magnet is "spin ice" \cite{Ramirez1994}. Naturally, the
question arises if and how itinerant electrons are affected by
such a frustrated lattice and what ground state is realized in
such systems. The pyrochlores constitute ideal systems for such a
study, since the network of the relevant metal atoms consists of
corner-sharing tetrahedra. Their electrical properties vary from
highly insulating to semiconducting to metallic, the pyrochlores
containing 5d transition metal ions being bad metals in general
\cite{SuArRa1983}. Surprisingly, superconductivity has been
reported for the first time in a pyrochlore for Cd$_2$Re$_2$O$_7$
\cite{HaMuTaSaYaHi2001,SaYoOhKaKaWaMaHaOn2001,JiHeMcAlDrMa2001},
and recently also in $\mathrm{KOs_2O_6}$ \cite{YoMuMaHi2004}, and
\fu\ \cite{HiYoMu2004}. Aside from the critical temperature, no
physical properties of \fu\ have yet been reported. Here we report
thermodynamic measurements on this compound.

The pyrochlore structure in general is of type
A$_2$B$_2$O$_6$O$'$, sometimes written as A$_2$B$_2$O$_7$, with
space-group F\,d\,$\bar{3}$\,m \cite{SuArRa1983}. The four
crystallographically inequivalent atoms in the face-centered cubic
unit cell A, B, O, and O$'$, occupy the 16d, 16c, 48f, and 8b
sites respectively. The compounds where A is a trivalent and B a
tetravalent cation (A$^{3+}$,B$^{4+}$), and also the
(A$^{2+}$,B$^{5+}$) combination, have been widely studied by
various workers. In the $\beta$-pyrochlore \fu, A is the
monovalent cation Rb, and B is the transition metal cation Os
octahedrally coordinated by O. It is derived from the parent
compound by replacing the O$'$ atoms by Rb atoms and leaving the
16d site empty (AB$_2$O$_6 \equiv \Box_2$B$_2$O$_6$A$'$, where
$\Box$ represents a vacancy). The Os atoms are coordinated by $6$
O atoms and these OsO$_6$ octahedra share corners. They have
edge-lengths of $2.67$ and $2.72\unit{\AA}$ respectively, and the
Os~-~O bonding distance is $1.91\unit{\AA}$. The electron donating
Rb atoms are coordinated by O with a Rb~-~O distance of
$3.13\unit{\AA}$. From X-ray measurements, we refine the 48f
oxygen $x$ parameter to $0.315(1)$. This results in O~-~Os~-~O
angles in the octahedra of $88.85\unit{^\circ}$ and
$91.15\unit{^\circ}$.

The B site may accommodate an ion carrying a localized magnetic
moment, which interacts antiferromagnetically with its nearest
neighbors. Since the B sublattice consists of interconnected
tetrahedra, we have to account for the 3-dimensional geometrical
frustration of the magnetic interactions, possibly leading to an
unconventional groundstate. The inset of Fig.~\ref{fig:XRD} shows
an Os atom and its relationship to its surrounding. Every Os atom
forms the shared corner of two Os tetrahedra in a network. The
exchange pathways to the nearest neighbors, leading by one of the
coordinating oxygen atoms, are shown by dashed lines. The pathways
form an Os~-~O~-~Os angle of $139.4\unit{^\circ}$, the direct
distance between the two Os being $3.58\unit{\AA}$. The Rb cations
are omited in this illustration for clarity. To our knowledge, the
B~-~O~-~B angle in \fu\ is surpassed only by one other pyrochlore,
the bulk superconductor Cd$_2$Re$_2$O$_7$. Together with the high
density of states at the Fermi level in this compound, we expect
the large B~-~O~-~B angle to play a crucial role in the tendency
to superconduct. It would therefore be interesting to know how it
compares to the corresponding angle in $\mathrm{KOs_2O_6}$.

Formally, the oxidation state of the osmium ion is either 5.5+ as
in Rb$^+$Os$^{5.5+}_2$O$^{2-}_6$ or $50\unit{\%}$ 5+ and
$50\unit{\%}$ 6+ as in Rb$^+$Os$^{5+}$Os$^{6+}$O$^{2-}_6$. Since
the ground state configuration of osmium is
[Xe]4f$^{14}$5d$^6$6s$^2$, one would expect the Os$^{5+}$ to be in
a spin $S=3/2$ state, while Os$^{6+}$ is in a spin $S=1$ state.
Band structure calculations for Cd$_2$(Os,Re)$_2$O$_7$
\cite{SiBlSchSo2002}, show that the crystal field well separates
the 5d e$_\mathrm{g}$ and t$_\mathrm{2g}$ bands from each other.
The 5d manifold is also well separated from the O~2p band.
Although the Os 16c site has trigonal symmetry, this part of the
crystal field is weak and there is no further splitting of the
t$_\mathrm{2g}$ manifold. However, the electronic situation in
\fu\ is expected to be quite different from
Cd$_2$(Os,Re)$_2$O$_7$: The 16d site is unoccupied and the Rb$^+$
cation is at the 8b site. Therefore, more detailed comparisons
require band-structure calculations specific for \fu.

Polycrystalline samples of \fu\ have been synthesized by a
procedure described in Ref.~\cite{YoMuMaHi2004}. A stoichiometric
amount of OsO$_2$ (Alfa Aesar, $99.99\unit{\%}$) and Rb$_2$O
(Aldrich, $99\unit{\%}$) was thoroughly mixed in an argon dry box
and pressed into a pellet. The pellet was put into a quartz tube
which was evacuated and sealed. This tube was heated up to
$600\unit{^{\circ}C}$ and kept at this temperature for
$24\unit{h}$. According to the X-ray diffraction analysis, the
resulting sample contained 2 phases: pyrochlore \fu\ and
RbOsO$_4$. RbOsO$_4$ can be removed by a $2\unit{h}$ treatment in
a $10\unit{\%}$ solution of HCl and subsequent washing in water
and drying at $100\unit{^{\circ}C}$. The X-ray diffraction pattern
of the purified sample is shown in Fig.~\ref{fig:XRD}, where all
reflections can be indexed on the basis of a pyrochlore unit cell.
The lattice parameter $a=10.1137(1)\unit{\AA}$ is slightly larger
than in $\mathrm{KOs_2O_6}$ ($a=10.099(1)\unit{\AA}$
\cite{YoMuMaHi2004}). The preliminary X-ray diffraction study
reveals that the Rb cations occupy the 8b site in the pyrochlore
lattice, as is the case for potassium in $\mathrm{KOs_2O_6}$
\cite{YoMuMaHi2004}. The experimental details of the synthesis and
purification of \fu\ are described elsewhere
\cite{KaZhBrBaKa2004}. $\mathrm{KOs_2O_6}$ and \fu\ are the only
reported osmium compounds in the form AB$_2$O$_6$
\cite{YoMuMaHi2004}, and both show bulk superconductivity.

\begin{figure}
  \includegraphics{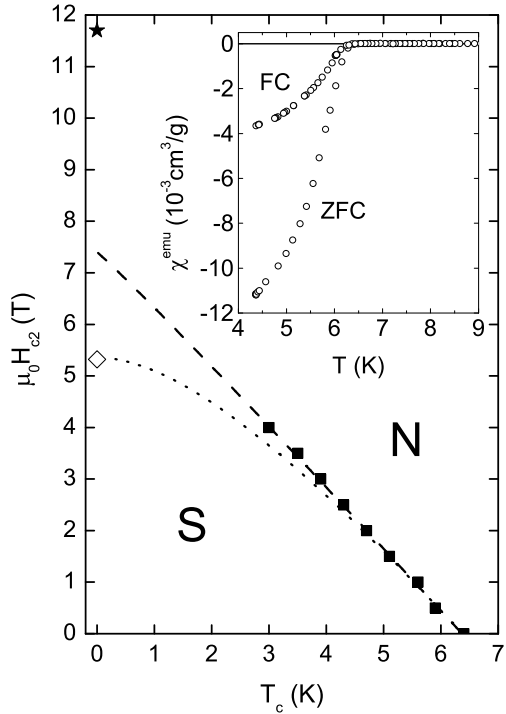}
  \caption{\label{fig:Hc2} Upper critical field $H_{\mathrm{c2}}$ extracted from specific heat
measurements. A power law fit $H_{\mathrm{c2}}(T) =
H_{\mathrm{c2}}(0) (1-(T/T_\mathrm{c})^n)$ (dashed line) gives an
exponent of $n=1.0 \pm 0.1$ and a critical temperature
$T_\mathrm{c} = (6.37 \pm 0.03)\unit{K}$. The initial slope of the
critical boundary at $T_\mathrm{c}$ is $-\mathrm{d}(\mu_0
H_{\mathrm{c2}})/\mathrm{d}T\vert_{T=T_\mathrm{c}} =
1.2\unit{T/K}$. The dotted line is a fit to the WHH formula in the
orbital limit ($\lambda_\mathrm{SO}=\infty$) using the calculated
Maki parameter $\alpha=0.64$. Also shown are the Pauli limiting
field $H_\mathrm{p0}$ ($\star$) and the orbital limiting field
$H_\mathrm{c2}^*$ ($\diamond$). The magnetic susceptibility at low
temperatures, depicted in the inset, shows the diamagnetic
transition into a bulk superconducting state. The magnetization
measurements have been performed at $H = 3.3 \unit{Oe}$ in
zero-field-cooled (ZFC) and field-cooled (FC) states. }
\end{figure}

The magnetic susceptibility at low temperatures, depicted in the
inset of Fig.~\ref{fig:Hc2}, shows the diamagnetic transition into
a bulk superconducting state. The measurements have been performed
at $H = 3.3 \unit{Oe}$ in zero-field-cooled and field-cooled
states.

\begin{figure}
  \includegraphics{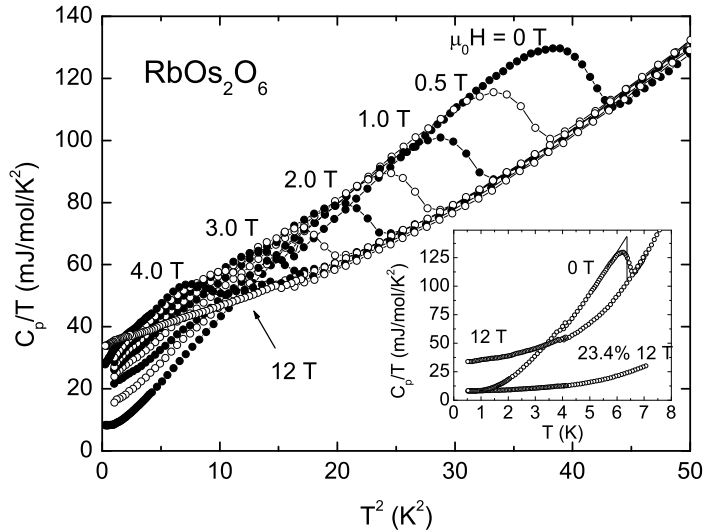}
  \caption{\label{fig:CdivTvsT2} Specific heat
$C_{\mathrm{p}}/T$ vs.~$T^2$ for magnetic fields from 0 to
$4\unit{T}$ in $0.5\unit{T}$ steps and $12\unit{T}$. From the
normal-state curve at $12\unit{T}$, we extract the electronic
specific heat coefficient $\gamma =
\lim_{T\to0\unit{K}}C_{\mathrm{p}}/T = (33.7 \pm
0.2)\unit{mJ/mol_{\mathrm{f.u.}}/K^2}$. The Debye temperature
$\Theta_\mathrm{D}(T)$ below $4\unit{K}$ is slightly temperature
dependent, with a value of $\Theta_\mathrm{D}(1\unit{K}) \approx
230\unit{K}$ and $\Theta_\mathrm{D}(4\unit{K}) \approx
240\unit{K}$. The inset shows the $0\unit{T}$ and normal state
specific heat on a linear $T$ scale. Also shown is $23.4\unit{\%}$
of the normal state $C_{\mathrm{p}}$, expected to be present in
the superconducting $C_{\mathrm{p}}$, and used for our analysis.
The normalized specific heat jump at $T_\mathrm{c}$, $\Delta
C_{\mathrm{p}}/(76.6\unit{\%}\gamma T_\mathrm{c})=1.38$, is close
to the weak-coupling BCS value of $1.43$.}
\end{figure}

Specific heat was measured in a physical properties measurement
apparatus using an adiabatic relaxation technique (Quantum Design,
PPMS). The sample was pressed into a cylindrical pellet of about
$20 \unit{mg}$. The applied magnetic field was perpendicular to
the cylinder axis. Figure \ref{fig:CdivTvsT2} shows the specific
heat $C_{\mathrm{p}}/T$ vs.~$T^2$ for magnetic fields from 0 to
$4\unit{T}$ in $0.5\unit{T}$ steps and also a normal state curve
measured at $12\unit{T}$.

From the normal-state curve at $12\unit{T}$, we extract the
electronic specific heat coefficient $\gamma =
\lim_{T\to0\unit{K}}C_{\mathrm{p}}/T$ by fitting the data below
$4\unit{K}$ to $C_{\mathrm{p}}/T = \gamma + \beta T^n$. From this
fit we get $\gamma =  (33.7 \pm
0.2)\unit{mJ/mol_{\mathrm{f.u.}}/K^2}$, $\beta = (1.42 \pm
0.1)\unit{mJ/mol_{\mathrm{f.u.}}/K^{2+n}}$, and $n=1.92 \pm 0.05$.
To our knowledge, this value of $\gamma$ is the largest reported
for a pyrochlore compound, and we expect it to play a crucial role
in the tendency for \fu\ to superconduct. Since the fit to the
low-temperature data yields $n < 2$, the Debye temperature
$\Theta_\mathrm{D}(T)$ in this range is slightly temperature
dependent, with a value of $\Theta_\mathrm{D}(1\unit{K}) \approx
230\unit{K}$ and $\Theta_\mathrm{D}(4\unit{K}) \approx
240\unit{K}$. The temperature dependence of $\Theta_\mathrm{D}(T)$
on a larger scale is typical for cubic metals \cite{Pa1958},
showing a broad minimum at about $11.5\unit{K}$ with a value of
$\Theta_\mathrm{D}(11.5 \unit{K}) \approx 190\unit{K}$. Towards
higher temperatures $\Theta_\mathrm{D}$ rises again to reach about
$225\unit{K}$ at $25\unit{K}$. Compared to other metallic
pyrochlores, which have a Debye temperature of typically $300$ to
$400\unit{K}$, $\Theta_\mathrm{D}$ in \fu\ is rather low,
indicative of lower frequency phonon modes.

\begin{figure}
  \includegraphics{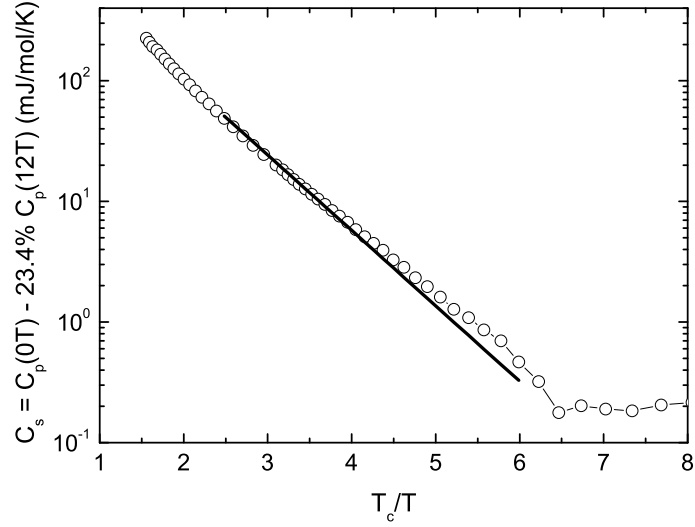}
  \caption{\label{fig:Cps_vs_Tinv} Superconducting specific heat
calculated by subtracting $23.4\unit{\%}$ of the normal state
specific heat at $12\unit{T}$, plotted on a semilogarithmic scale
vs.~$T_\mathrm{c}/T$. The corresponding superconducting volume
fraction is thus about $77\unit{\%}$, similar to the estimated
fraction in KOs$_2$O$_6$ \cite{YoMuMaHi2004}. The line indicates
the expected behavior from BCS assuming an isotropic gap: For
$2.5<T_\mathrm{c}/T<6$, the specific heat approximately follows an
exponential behavior $8.5 \, \gamma \, T_\mathrm{c} \exp(-1.44 \,
T_\mathrm{c}/T)$. Down to $1\unit{K}$, the superconducting
specific heat decreases in close quantitative agreement with this
conventional superconducting behavior.}
\end{figure}

In the superconducting state, a noticeable electronic density of
states at the Fermi level remains as $T \to 0\unit{K}$
(Fig.~\ref{fig:CdivTvsT2}). This might be interpreted as a sign of
an unconventional superconducting state with part of the Fermi
surface ungapped. Alternatively, if part of the sample does not
undergo a superconducting transition, then the measured curve is
the sum of the superconducting and normal contributions. Indeed,
such an analysis appears to yield consistent results, suggesting a
conventional BCS-like superconducting state. Following the second
interpretation, we proceed to further analyze $C_{\mathrm{p}}$
below $T_\mathrm{c}$. Starting with the $12\unit{T}$ data as the
normal state reference, we subtract $23.4\unit{\%}$ of it (shown
in the inset of Fig.~\ref{fig:CdivTvsT2}) from the $0\unit{T}$
set. Correspondingly, the superconducting volume fraction is about
$77\unit{\%}$, similar to the estimated fraction in KOs$_2$O$_6$
\cite{YoMuMaHi2004}. The resulting specific heat for the
superconducting part is shown in Fig.~\ref{fig:Cps_vs_Tinv} on a
semilogarithmic scale vs.~$T_\mathrm{c}/T$. The line indicates the
expected behavior from BCS assuming an isotropic gap
\cite{BaSch1961}: For $2.5<T_\mathrm{c}/T<6$, the specific heat
approximately follows an exponential behavior $8.5 \, \gamma \,
T_\mathrm{c} \exp(-1.44 \, T_\mathrm{c}/T)$. Down to $1\unit{K}$,
the specific heat in the superconducting state decreases in close
quantitative agreement with conventional superconducting behavior.
Below that, the subtraction of the two components provides
unreliable results, since the difference becomes exceedingly
small. Furthermore, impurities play an essential role at such low
temperatures, and it is thus difficult to draw conclusions from
the observed deviation. It will be worthwhile to elucidate the
microscopic origin of the roughly $20\unit{\%}$ ungapped fraction,
which is observed both in \fu\ and KOs$_2$O$_6$
\cite{YoMuMaHi2004}.

Additional support for the assumed two-component analysis comes
from the specific heat anomaly at $T_\mathrm{c}$: $\Delta
C_{\mathrm{p}}/T_\mathrm{c}$ is estimated to
$35.6\unit{mJ/mol_{\mathrm{f.u.}}/K^2}$ (inset of
Fig.~\ref{fig:CdivTvsT2}), resulting in a normalized ratio $\Delta
C_{\mathrm{p}}/(76.6\unit{\%}\gamma T_\mathrm{c})=1.38$. The
$76.6\unit{\%}$ in the denominator is the correction to $\Delta
C_{\mathrm{p}}$ due to the deviation of the superconducting volume
fraction from $100\unit{\%}$. This normalized specific heat jump
at $T_\mathrm{c}$ is indeed close to the weak-coupling BCS value
of $1.43$. Therefore, an analysis in terms of an about
$77\unit{\%}$ superconducting fraction appears reasonable, even as
it implies a very similar $\gamma$ for the superconducting and
non-superconducting parts of the sample. Since the X-ray
diffractogram (Fig.~\ref{fig:XRD}) shows the presence of only one
phase, the structure of the non-superconducting part has to be the
same as that of the superconducting one. Again, further analysis
is necessary to elucidate the microscopic origin of the ungapped
fraction.

Figure \ref{fig:Hc2} shows the upper critical field
$H_{\mathrm{c2}}$ extracted from specific heat measurements.
$T_\mathrm{c}$ has been chosen such that the entropy is balanced
above and below $T_\mathrm{c}$ from $C_{\mathrm{p}}/T$ vs.~$T$
data. The upper critical field of \fu\ lies below the
Pauli-limiting field $H_\mathrm{p0} = 1.84 \unit{T/K} \times
T_\mathrm{c} = 11.7 \unit{T}$ \cite{Cl1962}. This is in sharp
contrast to KOs$_2$O$_6$, where the critical field seems to exceed
the Pauli limit \cite{YoMuMaHi2004}. A power law fit
$H_{\mathrm{c2}}(T) = H_{\mathrm{c2}}(0) (1-(T/T_\mathrm{c})^n)$
(dashed line) gives an exponent of $n=1.0 \pm 0.1$ and a critical
temperature $T_\mathrm{c} = (6.37 \pm 0.03)\unit{K}$. Such a
linear behavior of $H_{\mathrm{c2}}$ is also observed in
Cd$_2$Re$_2$O$_7$ \cite{JiHeMcAlDrMa2001,SaYoOhKaKaWaMaHaOn2001}.
The initial slope of the critical boundary at $T_\mathrm{c}$ is
$-\mathrm{d}(\mu_0
H_{\mathrm{c2}})/\mathrm{d}T\vert_{T=T_\mathrm{c}} =
1.2\unit{T/K}$.

The upper critical field of a superconductor is determined by the
combined effect of an external magnetic field on the spin and
orbital degrees of freedom of the conduction electrons.
\citeauthor{WeHeHo1966}\ have worked out a theory for
$H_\mathrm{c2}$ which includes both spin and orbital paramagnetic
effects as well as nonmagnetic and spin-orbit scattering
\cite{WeHeHo1966}. In \fu, the upper critical field does not,
however, behave as predicted by the Werthamer-Helfand-Hohenberg
(WHH) formula in the dirty limit, as can be seen in
Fig~\ref{fig:Hc2}. The initial slope of the critical boundary at
$T_\mathrm{c}$ results in a Maki-parameter $\alpha = -0.52758
\unit{K/T} \times \mathrm{d} (\mu_0 H_\mathrm{c2})/\mathrm{d}T
\vert_{T=T_\mathrm{c}} = 0.64 \pm 0.1$ \cite{WeHeHo1966,Ma1964}.
Spin contributions to the energy balance of a superconductor
become important when $\alpha \gtrapprox 1$. Applying the WHH
model, we would therefore have to conclude that \fu\ is in the
orbital limit away from the Pauli-limiting regime with an orbital
critical field of $H_\mathrm{c2}^*(T=0) = \alpha H_\mathrm{p0} /
\sqrt{2} = 5.3\unit{T}$. This field might be a good extrapolation
of the critical field to $0\unit{K}$ judging from
Fig.~\ref{fig:Hc2}. However, calculations based on a realistic
Fermi surface are needed to further clarify the detailed behavior
of $H_\mathrm{c2}$. Using a reasonably extrapolated
$H_\mathrm{c2}$ of $6\unit{T}$, we calculate the Ginzburg-Landau
coherence length at $0\unit{K}$ of $\xi = \sqrt{\Phi_0/(2 \pi
H_\mathrm{c2})} \approx 74\unit{\AA}$, where $\Phi_0$ is the
magnetic flux quantum.

We have studied the basic thermodynamic properties of the new
pyrochlore superconductor \fu\, with a critical temperature of
$6.4\unit{K}$. The electronic density of states at the Fermi level
in the normal state results in $\gamma =  (33.7 \pm
0.2)\unit{mJ/mol_{\mathrm{f.u.}}/K^2}$, the highest reported value
for a pyrochlore. Together with the high $\gamma$, we expect the
large Os~-~O~-~Os angle of $139.4\unit{^\circ}$ to play a crucial
role in the tendency towards superconductivity. The inherent
geometrical frustration of the three dimensional tetrahedral Os
sublattice attributes further importance to this new
superconductor. We find that a careful analysis of the specific
heat data, taking into account the normal-state volume fraction,
implies conventional BCS-type behavior. Although the compound
could be expected to be only marginally different from the related
compound $\mathrm{KOs_2O_6}$, a significant difference in the
magnitude of the upper critical field $H_{\mathrm{c2}}$ is
observed. Considering the potential for competing electronic
groundstates in these compounds, the pyrochlore type
superconductors may be of fundamental importance to address
unsolved issues.

This study was partly supported by the Swiss National Science
Foundation.


\newpage


\begin{figure}
  \caption{\label{fig:XRD} The powder X-ray diffraction pattern of
the purified sample, where all reflections can be indexed on the
basis of a pyrochlore face-centered unit cell with lattice
parameter $a=10.1137(1)\unit{\AA}$. The inset shows an Os atom and
its relationship to its surrounding. Every Os atom forms the
shared corner of two Os tetrahedra in a network. The exchange
pathways to the nearest neighbors, leading by one of the
coordinating oxygen atoms, are shown by dashed lines. The Rb
cations are omited in this illustration for clarity.}
\end{figure}

\newpage

\newpage

\newpage

\end{document}